\documentclass[conference,a4paper]{IEEEtran}
\usepackage[noadjust]{cite}
\usepackage{amsmath}
\usepackage{amsfonts}
\usepackage{graphicx}
\usepackage{float}
\usepackage{subfigure}
\usepackage{epsfig}
\usepackage{amsmath}
\usepackage{theorem}
\usepackage{amsmath}
\usepackage{mathrsfs}
\usepackage{amsfonts}
\usepackage{amssymb}
\usepackage{mathrsfs}
\usepackage{euscript}
\usepackage{multirow}
\usepackage{cite}
\usepackage{url}
\usepackage{color}
\usepackage{algorithm}
\usepackage{algorithmic}
\usepackage{bm}
\usepackage{flushend}
\def\BibTeX{{\rm B\kern-.05em{\sc i\kern-.025em b}\kern-.08em
    T\kern-.1667em\lower.7ex\hbox{E}\kern-.125emX}}
\pdfhorigin=\dimexpr1in+(210mm-8.5in)/2\relax
\pdfvorigin=\dimexpr1in+(297mm-11in)/2\relax

\newtheorem{proposition}{Proposition}

\begin{document}
\title{A Kronecker-Based Sparse Compressive Sensing Matrix for Millimeter Wave Beam Alignment}
%\author{Author~1, %,~\IEEEmembership{Student Member,~IEEE,}
 %       Author~2%,~\IEEEmembership{Fellow,~OSA,}
\author{Erfan Khordad, Iain B. Collings, Stephen V. Hanly \\~Department of Engineering, Macquarie University, Sydney, Australia\\
Emails: erfan.khordad@hdr.mq.edu.au, iain.collings@mq.edu.au, stephen.hanly@mq.edu.au}
%\markboth{IEEE COMMUNICATIONS LETTERS,~Vol.~, No.~, ...~2018}%$^{\dag}$
%{Shell \MakeLowercase{\textit{et al.}}: Bare Demo of IEEEtran.cls for Journals}
\IEEEoverridecommandlockouts
%\IEEEpubid{\makebox[\columnwidth]{978-1-5386-5541-2/18/\$31.00~\copyright2019 IEEE \hfill} \hspace{\columnsep}\makebox[\columnwidth]{ }}
\maketitle
\IEEEpubidadjcol
\begin{abstract} 
Millimeter wave beam alignment~(BA) is a challenging problem especially for large number of antennas. Compressed sensing~(CS) tools have been exploited due to the sparse nature of such channels. This paper presents a novel deterministic CS approach for BA. Our proposed sensing matrix which has a Kronecker-based structure is sparse, which means it is computationally efficient. We show that our proposed sensing matrix satisfies the restricted isometry property~(RIP) condition, which guarantees the reconstruction of the sparse vector. Our approach outperforms existing random beamforming techniques in practical low signal to noise ratio~(SNR) scenarios. 
\end{abstract}

\IEEEpeerreviewmaketitle
\section{Introduction}
One of the potential key enablers of 5G is employing mmWaves due to the large bandwidth available at these wavelengths. The challenges regarding mmWaves mostly pertain to their adverse propagation characteristics~\cite{WW5GBB}.  High antenna gain using directional beamforming~(BF) of large antenna arrays can be a potential solution for this issue~\cite{Mwbaaetf5gcctfapr}. This is feasible owing to the small wavelength of mmWaves, and it is done by compacting a large number of antenna elements in a small size.

It has been shown that scatterers in mmWave propagation channels follow a sparse pattern~\cite{cCsanatEsmch}\cite{D60Gcfmgpswfip}. In other words, there are only a few strong propagation paths in the mmWave channel. It is critical that base stations~(BSs) and user equipments~(UEs) find the strong propagation paths in the BA process so as to align their beams in those directions. The problem of determining the best beam directions in terms of SNR for the connection between transceivers is called Beam Alignment (BA)~\cite{aSaSrbatfmws}.

A simple approach for BA is exhaustive search probing all possible combinations of generated beams by transceivers. This method is particularly favorable since narrow beams can be used to obtain high SNR; however, it yields a large training overhead. Hierarchical search is another approach reducing the total number of measurements. In this approach, transceivers first use wider beams and then based on the feedback exchanged between them, they refine the beams to finally find the best beam pair with the desired resolution~\cite{Mmwscbsdbaasdch}. In multiuser scenarios, hierarchical search might not be an efficient approach since it requires to be carried out for every single user, and therefore, the training overhead depends linearly upon the number of users.

Employing the CS tool is a new approach to BA in order to exploit the sparse nature of the mmWave channels. Non-adaptive CS approaches are specifically useful for multiuser scenarios because each UE can
estimate its own channel separately, which means that growing the number of users leads to no extra
training overhead and all UEs can estimate their respective channels concurrently~\cite{cEviaOMpfhmsimwcs}\cite{csbmmmwshmman}. In addition, deploying the CS tool,  according to CS fundamentals~\cite{cS2006DLDonoho}, can significantly reduce the number of measurements required for the estimation in case where the unknown vector is sparse.
 
The structure of the sensing matrix employed in CS has a key role in successful recovery. In fact, properties of the sensing matrix determine the possibility of perfect recovery. It has been shown that random sensing matrices constructed based on Gaussian and Bernoulli distributions satisfy the RIP condition with high probability~\cite{asimpleproofofRIP}. This means using a random sensing matrix guarantees the sparse recovery with high probability. Consequently, using random BF vectors which leads to a random sensing matrix is a favorable method for mmWave BA.

In~\cite{csbmmmwshmman} the components of BF vectors are generated randomly, leading to a complex random sensing matrix. They quantized the angles of arrival and departure~(AoA/AoDs), using the idea of virtual channel representation~\cite{cCsanatEsmch}\cite{Dmafadingch}. The sparse channel is reconstructed by orthogonal matching pursuit~(OMP) algorithm. One of the important steps in the OMP algorithm is solving a least square problem~\cite{cEviaOMpfhmsimwcs} which involves taking the inverse of a matrix containing columns of the sensing matrix whose component are non-zero complex values in~\cite{csbmmmwshmman}. When the sparsity level is large, the inverse matrix is also large, and as such the computational complexity of OMP is high.

A structured random CS method is presented in~\cite{strCCSfmmwlsas}. A column of the DFT matrix is randomly selected as a beamformer to generate a beam in the first stage and the beam is spread over the entire angular range using a unimodular sequence in the second stage. The sparse formulation in~\cite{strCCSfmmwlsas} is based on a circulant convolution between the virtual channel representation of the channel matrix and circulant matrices, spreading the information of the virtual channel representation uniformly in the angle domain. Since the OMP algorithm is deployed to recover the sparse channel and all components of the proposed sensing matrix are non-zero complex values, the proposed method in~\cite{strCCSfmmwlsas} suffers from large computational complexity as was the case for the approach in~\cite{csbmmmwshmman}.

Second order statistics of the channel is used in~\cite{aSaSrbatfmws} in order to propose a robust BA method against the
significant variation of the channel in mmWave systems. It is assumed in~\cite{aSaSrbatfmws} that AoA/AoDs do not significantly vary over the BA period. The BF vectors at the BS and the UEs are random linear combinations of the columns of the discrete Fourier transform (DFT) basis in~\cite{aSaSrbatfmws}. This leads to a CS formulation with a random sensing matrix including only zeros and ones. The random sensing matrix employed in~\cite{aSaSrbatfmws} works well for CS methods with high probability; however, it is not guaranteed that a specific realization of the sensing matrix resulting from random BF codebooks always works~\cite{DCobbatcsmAmini}.

In this paper, we propose a deterministic sensing matrix for the BA problem. Our proposed deterministic sensing matrix which has a Kronecker-based structure is inherently sparse, which leads to a more computationally efficient reconstruction algorithm compared to the sensing matrices employed in~\cite{csbmmmwshmman} and~\cite{strCCSfmmwlsas}. This contributes to faster measurement process and lower amount of computational burden at the UE's receiver. Also, to construct the proposed deterministic sensing matrix, the UE needs to have access to only a few parameters sent by the BS or stored in the UE's memory, which means our approach results in significant overhead reduction. We show that our proposed sensing matrix satisfies the RIP and mutual incoherence property, guaranteeing the sparse recovery. In addition, we design the BF vectors for the BA process based on our proposed deterministic sensing matrix.

\section{System Model}\label{system}
We consider a mmWave wireless system comprising a BS with $N_T$ antennas and a generic UE with $N_R$ antennas where the BS and the UE both are equipped with uniform linear arrays~(ULAs). The space between antenna elements in the arrays is $d=\frac{\lambda}{2}$, where $\lambda$ is the wavelength and it is calculated by $\lambda=\frac{c_0}{f_0}$, and $c_0$ and $f_0$ are the speed of light and the carrier frequency respectively. We further assume that phase shifting as well as the amplitude control can be performed in the analog domain. This is a practically feasible assumption for mmWave systems as it has been shown in the literature~\cite{XiaoshenAnalogAmplitude}\cite{Majidzadeh7980696}. Assuming $\theta_l \in [-\frac{\pi}{2},\frac{\pi}{2}]$ and $\phi_l \in [-\frac{\pi}{2},\frac{\pi}{2}]$ respectively the AoD and AoA of the $l$th propagation path between the BS and UE, the array response vectors are given by
\begin{equation}
{\mathbf{a}}(\theta_l)=\frac{1}{\sqrt{N_T}}[1, e^{j\pi \sin (\theta_l)},...,e^{j (N_T-1)\pi \sin (\theta_l)}]^T,
\end{equation}
and

\begin{equation}
{\mathbf{b}}(\phi_l)=\frac{1}{\sqrt{N_R}}[1, e^{j\pi \sin (\phi_l)},...,e^{j (N_R-1)\pi \sin (\phi_l)}]^T.
\end{equation}
We also assume that AoDs and AoAs of the propagation paths have uniform distribution within the angular range $[-\frac{\pi}{2},\frac{\pi}{2}]$.

Since a small number of clusters contributes to the propagation paths in the mmWave channels~\cite{Dmafadingch}\cite{MMWcMacce}, we use the clustered physical channel model as follows:
 \begin{equation}\label{Physical_Ch_Model}
{{\mathbf H}}=\sqrt{\frac{N_T N_R}{L}}\sum\limits_{l = 1}^L {{\alpha _l}{{\mathbf b}}(\phi
 _l){{\ }}{{{\mathbf a}}^H}(\theta _l)},
 \end{equation}
where we assume that the physical channel includes $L$ clusters of scatterers each of which creates a propagation path and $L \ll \max\{N_T , N_R\}$~\cite{aSaSrbatfmws}. Also, $\alpha_l \sim \mathcal{CN}(0,\,\sigma^{2}_{\alpha_l})$ is the complex channel gain of the $l$th propagation path. In addition, we assume that all path gains, i.e., $\alpha _l$, are constant during the beam alignment~(BA) procedure. This is relevant to dense mmWave networks~\cite{CIDmmwavecn}\cite{csbmmmwshmman}.

In~(\ref{Physical_Ch_Model}), the AoAs and AoDs have continuous values. To have a tractable channel model, we approximate the channel model in~(\ref{Physical_Ch_Model}) with a discrete representation using the idea of a virtual channel model (or beamspace representation)~\cite{Dmafadingch}. To do so, we use the following expressions to quantize the AoAs and AoDs:
\begin{equation}\label{quantizedAnglesTheta}
\sin(\theta_{c_1}^q) = \frac{2(c_1-1)}{N_T}-1;~c_1 = 1,2,...,N_T,
\end{equation}
\begin{equation}\label{quantizedAnglesFi}
\sin(\phi_{c_2}^q) = \frac{2(c_2-1)}{N_R}-1;~c_2 = 1,2,...,N_R,
\end{equation}
where $\theta_{c_1}^q$ and $\phi_{c_2}^q$ indicate the quantized angles. Since ULAs are employed at the BS and UE, the array response vectors corresponding to all $\theta_{c_1}^q$ and all $\phi_{c_2}^q$ form unitary DFT matrices as follows:
\begin{equation}\label{DFTByArrayResponseVectNT}
{\mathbf{F}}_{N_T} = [{\mathbf{a}}(\theta_{1}^q), {\mathbf{a}}(\theta_{2}^q), . . . , {\mathbf{a}}(\theta_{N_T}^q)],
\end{equation}
and
\begin{equation}\label{DFTByArrayResponseVectNR}
{\mathbf{F}}_{N_R} = [{\mathbf{b}}(\phi_{1}^q), {\mathbf{b}}(\phi_{2}^q), . . . , {\mathbf{b}}(\phi_{N_R}^q)],
\end{equation}
Now, using the DFT matrices, we can represent the channel model by
\begin{equation}\label{VchRep}
{\mathbf H} = \mathbf F_{N_R} \mathbf H_v  \mathbf F_{N_T}^H,
\end{equation}
where $\mathbf H_v$ is the virtual channel representation which is a sparse matrix with $L$ components having significant nonzero values corresponding to the AoAs and AoDs of the propagation paths.

In the training process, the BS transmits pilot signals $x_t$ using unit-norm transmit BF vectors $\mathbf w_t \in \mathbb{C}^{N_T \times 1}$, where $t$ denotes the $t$-th measurement. Then, the UE applies its unit-norm receive BF vectors $\mathbf g_t \in \mathbb{C}^{N_R \times 1}$ to make the $t$-th measurement which is given by
\begin{equation}\label{measurementEquation}
y_t = \mathbf g_t^H \mathbf H \mathbf w_t x_t + \mathbf g_t^H \mathbf n_t,
\end{equation}
where $\mathbf n_t \sim \mathcal{CN}(\mathbf 0,\,\sigma_n^{2}\mathbf{I})$ is the noise vector. Without loss of generality, we assume that $x_t=\sqrt{P}$, where $P$ is the average received power of the pilot signals.

By applying the vectorization identity $vec(\mathbf{ABC}) = (\mathbf C^T \otimes \mathbf A)vec(\mathbf B)$ to both sides of (\ref{measurementEquation}) where $\otimes$ indicates the Kronecker product and defining $\mathbf h = vec(\mathbf H)$, we can write
\begin{equation}\label{AfterVectorizationEq}
y_t = \sqrt{P} ( \mathbf w_t^T \otimes \mathbf g_t^H ) \mathbf h  +  \mathbf g_t^H \mathbf n_t.
\end{equation}
Referring to (\ref{VchRep}), we can obtain $\mathbf h$ in terms of the virtual channel representation as follows:
\begin{equation}\label{VectorizationVirtualchModel}
\mathbf h = vec(\mathbf H) = (\mathbf F_{N_T}^* \otimes \mathbf F_{N_R}) \mathbf h_v,
\end{equation}
where $\mathbf h_v = vec(\mathbf H_v)$.

In this paper, we use the linear combinations of the columns of the DFT matrices to design the beam patterns as described in~\cite{aSaSrbatfmws}. The vectors $\mathbf w_t^{(1)} \in \{0,1\}^{N_T \times 1}$ and $\mathbf g_t^{(1)} \in \{0,1\}^{N_R \times 1}$ select the columns of the DFT matrices $\mathbf F_{N_T}$ and $\mathbf F_{N_R}$ respectively. Therefore, the transmit and receive BF vectors can respectively be expressed as
\begin{equation}\label{multifingerTransitEqW}
\mathbf w_t = \mathbf F_{N_T}\frac{\mathbf w_t^{(1)}}{\sqrt{z_1}},~~~\mathbf g_t = \mathbf F_{N_R}\frac{\mathbf g_t^{(1)}}{\sqrt{z_2}},
\end{equation}
where $z_1$ and $z_2$ indicate the number of ones in ${\mathbf w_t^{(1)}}$ and ${\mathbf g_t^{(1)}}$. Note that each component of the vectors ${\mathbf w_t^{(1)}}$ or ${\mathbf g_t^{(1)}}$ is related to one quantized angle. If a component of these vectors equals one, it indicates that the antenna elements generate a narrow beam aligned with the corresponding quantized angle to that component. In fact, the ones in the vectors ${\mathbf w_t^{(1)}}$ or ${\mathbf g_t^{(1)}}$ can be thought of as switching on the corresponding narrow beams and the zeros are for switching off the corresponding narrow beams. Multiple ones in the vectors ${\mathbf w_t^{(1)}}$ or ${\mathbf g_t^{(1)}}$ result in multiple narrow beams.

Using (\ref{AfterVectorizationEq}), (\ref{VectorizationVirtualchModel}) and (\ref{multifingerTransitEqW}), we can write
\begin{equation}\label{vectoronessubstitution}
y_t = \sqrt{\frac{P}{z_1 z_2}} ( \mathbf w_t^{{(1)}^T} \mathbf F_{N_T}^T \otimes \mathbf g_t^{{(1)}^H} \mathbf F_{N_R}^H ) (\mathbf F_{N_T}^* \otimes \mathbf F_{N_R}) \mathbf h_v  +  \tilde{n}_t.
\end{equation}
Also, using the identity $(\mathbf A \otimes \mathbf B)(\mathbf C \otimes \mathbf D) = \mathbf {AC} \otimes \mathbf {BD}$ and assuming $P=1$, (\ref{vectoronessubstitution}) can be rewritten as
\begin{equation}\label{finalmeasurementsEq}
y_t = {\frac{1}{\sqrt{z_1 z_2}}} (\mathbf w_t^{{(1)}^T} \otimes \mathbf g_t^{{(1)}^T}) \mathbf h_v + \tilde{n}_t,
\end{equation}
where $\tilde{n_t} \sim \mathcal{CN}(0,\,\sigma_n^{2})$, and as $\mathbf g_t^{(1)}$ is a real-valued vector (it contains only ones and zeros), $\mathbf g_t^{{(1)}^H}$ has been substituted with $\mathbf g_t^{{(1)}^T}$.

The measurement vector $\mathbf y$ can be formed at the UE by stacking all the measurements and it can be written as follows:
\begin{equation}\label{StackingAllMeasurementsEq}
\mathbf y = \mathbf S \mathbf h_v + \tilde{\mathbf n},
\end{equation}
where the sensing matrix $\mathbf S$ is given by
\begin{equation}\label{SensingMatrixEq}
\mathbf S = {\frac{1}{\sqrt{z_1 z_2}}}\mathbf S_b~~;
\mathbf S_b =  \begin{pmatrix}
   {\mathbf w^{(1)}_1}^T \otimes { \mathbf g^{(1)}_1}^T  \\
  {\mathbf w^{(1)}_2}^T \otimes { \mathbf g^{(1)}_2}^T  \\
\vdots\\
  {\mathbf w^{(1)}_k}^T \otimes { \mathbf g^{(1)}_k}^T
 \end{pmatrix}.
\end{equation}
Note that $k$ denotes the number of measurements, $\tilde{\mathbf n}$ in (\ref{StackingAllMeasurementsEq}) is the noise vector and $\mathbf S_b$ is a binary matrix.

Since $\mathbf h_v $ has a sparse structure, a proper sensing matrix $\mathbf S$ should be employed for a good reconstruction of $\mathbf h_v $. As seen, the structure of $\mathbf S$ depends on the vectors $\mathbf w_t^{(1)}$ and $\mathbf g_t^{(1)}$ which make the transmit and receive BF vectors. In fact, the vectors $\mathbf w_t^{(1)}$ and $\mathbf g_t^{(1)}$ show how the measurements are made in the angular domain.
\section{RIP and Incoherence Property}
In this section, we define the RIP and incoherence property, which will be used in our proposed scheme. Let $\mathbf{s}_{n\times1}$ be an $L$-sparse vector. The noiseless CS problem can be stated as $\mathbf{y}=\mathbf{\Theta}\mathbf{s}$, where $\mathbf{y}_{m\times1}$ and $\mathbf{\Theta}_{m\times n}$ indicate the measurement vector and the sensing matrix respectively. The restricted isometry property (RIP) is a sufficient condition for stable reconstruction~\cite{dblp}\cite{ssrfiaims}. The sensing matrix $\mathbf{\Theta}$ satisfies the RIP of order $L$ if for all $L$-sparse vectors $\mathbf{u}$ and a constant $0< \delta_L < 1$, the following condition holds:
\begin{equation}\label{RIP_condition}
1-\delta_L \leq \frac{\|\mathbf{\Theta}\mathbf{u}\|_2}{\|\mathbf{u}\|_2} \leq 1+\delta_L.
\end{equation}
Given $\mathbf{\Theta}$, $L$ and $\delta_L$, it is an arduous task to verify RIP~\cite{ctRIPih}. An easier condition to verify is the mutual incoherence property. To measure the mutual coherence of $\mathbf{\Theta}$ the following expression is used
\begin{equation}\label{Incoherence_condition}
\mu(\mathbf{\Theta}) = \max_{i\neq j}\frac{|\langle \bm{\theta}_i , \bm\theta_j \rangle|}{\|\bm\theta_i\|_2 \|\bm\theta_j\|_2},
\end{equation}
where $\bm\theta_i$ are the columns of $\mathbf{\Theta}$. The value of $\mu$ is bounded between the Welch bound $\sqrt{\frac{n-m}{m(n-1)}}$ and one, and a small value of $\mu$ is desirable~\cite{asRocsciaa}.
\section{Proposed Kronecker-based Spare Sensing Matrix}\label{problem}
In this section, we propose a deterministic sensing matrix, and based on the proposed deterministic sensing matrix, we design the structure of the BF vectors for the BA process. We construct the deterministic sensing matrix by performing a Kronecker product between two existing sensing matrices.

Since we intend to construct a deterministic sensing matrix for the sparse formulation~(\ref{StackingAllMeasurementsEq}), we should design deterministic BF codebooks for the BS and UE. As we showed earlier, each measurement is made based on the Kronecker product of the two vectors ${\mathbf w_t^{(1)}}^T$ and ${\mathbf g_t^{(1)}}^T$. Therefore, if we can obtain the sensing matrix by performing a Kronecker product between two matrices $\mathbf X_1$ and $\mathbf X_2$, the rows of $\mathbf X_1$ and $\mathbf X_2$ (after applying a transpose operation) can be used respectively as $\mathbf w_t^{(1)}$ at the BS and $\mathbf g_t^{(1)}$ at the UE to generate the BF vectors. The following proposition shows that indeed this is possible when $\mathbf X_1$ and $\mathbf X_2$ themselves are sensing matrices.
\begin{proposition}\label{proposotionKron}~\cite{JOKAR20092437}
If $\mu (\mathbf X_1)$ and $\mu (\mathbf X_2)$ are the mutual coherence of $\mathbf X_1$ and $\mathbf X_2$ respectively, and $\mu (\mathbf W)$ is the mutual coherence of $\mathbf W$ where $\mathbf W= \mathbf X_1 \otimes \mathbf X_2$, we have $\mu (\mathbf W) = \max \{\mu (\mathbf X_1) , \mu (\mathbf X_2) \}$.
\end{proposition}
DeVore in~\cite{DEVORE2007918} designed $p^2 \times p^{r+1}$ binary deterministic sensing matrices with mutual coherence $\frac{r}{p}$, where $p$ is a prime power and $1\leq r < p$. Also, the DeVore's matrix with normalized columns satisfies RIP of order $L<\frac{p}{r} + 1$ with RIP constant $\delta_L = (L-1)\frac{r}{p}$. Employing the DeVore's approach, we construct two binary deterministic sensing matrices $\mathbf U_b$ and $\mathbf V_b$ respectively with dimensions $p_1^2 \times p_1^{r_1+1}$ and $p_2^2 \times p_2^{r_2+1}$. Now, using preposition \ref{proposotionKron}, $\mathbf S_b$ in (\ref{SensingMatrixEq}) can be constructed as $\mathbf S_b = \mathbf U_{b_{p_1^2 \times p_1^{r_1+1}}} \otimes \mathbf V_{b_{p_2^2 \times p_2^{r_2+1}}}$ which has $\mu (\mathbf S_b) = \max \{\frac{r_1}{p_1} , \frac{r_2}{p_2} \}$. The number of components of $\mathbf h_v$ is $N_TN_R$, so we assume that $N_T$ and $N_R$ are equal to the number of columns of $\mathbf U_b$ and $\mathbf V_b$ respectively, i.e., $N_T=p_1^{r_1+1}$ and $N_R=p_2^{r_2+1}$. In other words, the number of antennas at the BS and UE are assumed to be prime powers.

In the DeVore’s matrix, the number of ones in each row is a constant value. Assuming that there are $c_U$ ones and $c_V$ ones in each row of $\mathbf U_b$ and in each row of $\mathbf V_b$ respectively, by multiplying $\mathbf U_b$ and $\mathbf V_b$ by normalization factors $\frac{1}{\sqrt{c_U}}$ and $\frac{1}{\sqrt{c_U}}$ respectively, $\mathbf U$ and $\mathbf V$ which are the row-normalized version of $\mathbf U_b$ and $\mathbf V_b$ are obtained. In fact, referring to~(\ref{multifingerTransitEqW}) and (\ref{SensingMatrixEq}), we assume that $z_1=c_U$ and $z_2=c_V$, which makes $\mathbf S$ in (\ref{SensingMatrixEq}) row-normalized. Also, as we need unit-norm BF vectors in the BA process, we define $\tilde{\mathbf w}_t = \frac{\mathbf w_t^{(1)}}{\sqrt{z_1}}$ and $\tilde{\mathbf g}_t = \frac{\mathbf g_t^{(1)}}{\sqrt{z_2}}$ and we use each row of $\mathbf U$ as the vectors $\tilde{\mathbf w}_t$ and each row of $\mathbf V$ as the vectors $\tilde{\mathbf g}_t$.

Because of the structure of the Kronecker product, each vector $\tilde{\mathbf w}_t$, is repeated $p_2^2$ times for all the different vectors $\tilde{\mathbf g}_t$. This means that the BS repeats the same transmit BF vector (or the same beam pattern) for $p_2^2$ times while the UE probes the channel using its all possible beam patterns. Then, the BS uses its second beam pattern and repeats it for $p_2^2$ times while the UE again probes the channel using its all possible beam patterns. This process continues until all possible combinations of the BS's beam patterns and UE's beam patterns are used to probe the channel.

\section{RIP Condition for the proposed appraoch}
Our design results in a row-normalized $\mathbf S$. The CS formulation in (\ref{StackingAllMeasurementsEq}) can be converted to an equivalent CS formulation with a column-normalized sensing matrix. To do so, we need to remark that each column of the DeVore's sensing matrix has $p$ ones~\cite{DEVORE2007918}. Therefore, each column of $\mathbf U_b$ and $\mathbf V_b$ have $p_1$ and $p_2$ ones respectively. Now, we can rewrite (\ref{StackingAllMeasurementsEq}) as follows:
\begin{equation}\label{StackingAllMea2}
\mathbf y = \mathbf S_{C}\tilde{ \mathbf h}_v + \tilde{\mathbf n},
\end{equation}
where $\mathbf S_{C}=\sqrt{\frac{z_1z_2 }{p_1p_2}}\mathbf S$ is column-normalized and $\tilde{ \mathbf h}_v = \sqrt{\frac{p_1p_2}{z_1z_2}}\mathbf h_v$.
Note that in (\ref{StackingAllMea2}) the measurement vector $\mathbf y$ and the noise vector $\tilde{\mathbf n}$ are the same as those of (\ref{StackingAllMeasurementsEq}). Consequently, the UE, after making all the measurements, can use (\ref{StackingAllMea2}) for the sparse recovery process.

The sensing matrix $\mathbf S_{C}$ is column-normalized and its mutual coherence is $\mu (\mathbf S_C) = \max \{\frac{r_1}{p_1} , \frac{r_2}{p_2} \}$; therefore, according to the following proposition it satisfies the RIP.
\begin{proposition}\label{JeanBourgainProposition}~\cite{ExoripandrpDuke}
The sensing matrix $\mathbf{\Theta}$ with unit-norm columns and the coherence parameter $\mu_N$ satisfies RIP of order $L$ with constant $\delta_L = (L-1)\mu_N$
\end{proposition}

\section{Simulation Results}\label{simulation}
In this section, we compare the performance of our proposed approach with the random BF design proposed in~\cite{aSaSrbatfmws}. Based on the proposed BF design in~\cite{aSaSrbatfmws}, the number of ones in $\mathbf w_t^{(1)}$ and $\mathbf g_t^{(1)}$ are constant but their positions are randomly permuted. We call this method random permutation and we denote it by the abbreviation RdPerm. In addition, we call our proposed approach matrix-by-matrix Kronecker product~(MbMKP) in the simulations results. In our proposed approach, the number of ones in $\mathbf w_t^{(1)}$ and $\mathbf g_t^{(1)}$ are also constant, but the positions of ones are fixed because we have designed the BF vectors based on our proposed deterministic sensing matrix.

The BS and UE are equipped with $N=N_T=N_R=\{27,64\}$ antennas, i.e., the pairs of $\{p_1=p_2=3, r_1=r_2=2\}$ and $\{p_1=p_2=4, r_1=r_2=2\}$ are used to construct $\mathbf U_b$ and $\mathbf V_b$. We assume one propagation path in the mmWave channel ($L=1$), and we set $\sigma^{2}_{\alpha_1}=1$. Also, to estimate the index of the strongest component in $\mathbf h_v$, we use the OMP algorithm. The SNR in our simulations is defined as $\text{SNR}=\frac{P}{\sigma^2_n}$.

In mmWave systems, typically the SNR in the beam alignment process is very low~\cite{aSaSrbatfmws}. Thus, it is reasonable that first the directions of the propagations path between the BS and UE are found, and then the path gains are estimated when the beams are aligned in those direction. Thus, we use the probability of correct alignment~(PCA) as a performance metric. Correct alignment means that the directions of the propagation paths are found correctly, which is equivalent to the probability of correctly finding the index of the strongest element in the sparse vector $\mathbf h_v$.

We use the SNR after BF~(SNR$_\text{AB}$) as another performance metric. After the beam alignment process, the BS and UE can align their narrow beams in the direction of the propagation path. To do so, the BS and UE need to know the indexes of the value one in $\mathbf w_t^{(1)}$ and $\mathbf g_t^{(1)}$ respectively. If we denote by $\varepsilon$ the index of a nonzero element in $\mathbf h_v$, the indexes of the value one in $\mathbf w_t^{(1)}$ and $\mathbf g_t^{(1)}$ are calculated by $\varepsilon_{\mathbf w} = \lfloor (\varepsilon-1)/ N_R\rfloor+1$ and $\varepsilon_{\mathbf g} =  \big((\varepsilon-1) \mod N_R\big)+1$ respectively. Then, the SNR$_\text{AB}$ is calculated as follows:
\begin{equation}\label{SNR_ABeq}
\text{SNR}_\text{AB} = \frac{{|\mathbf g_t^{(1)}}^H \mathbf H_v \mathbf w_t^{(1)}|^2}{\sigma^2_n}
\end{equation}

\begin{figure}[t]
\centering
 \includegraphics[scale=.6]{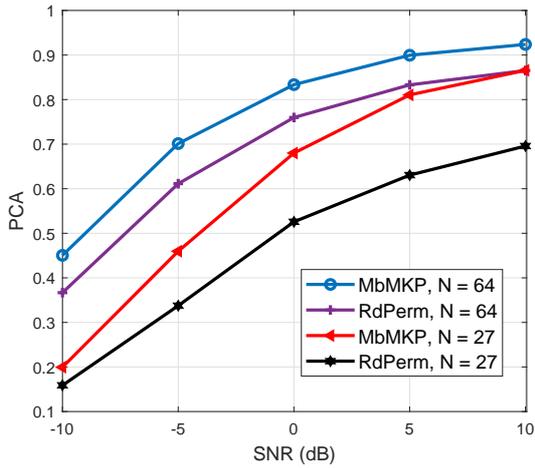}
  \caption{PCA vs SNR for $L=1$.}\label{OneCluster64ant}
\end{figure}

In Fig.~\ref{OneCluster64ant}, we compare the performance of our proposed approach with RdPerm in terms of PCA. For the scenarios with $N=27$ and $N=64$ the number of measurements are respectively $m=p_1^2p_2^2=81$ and $m=p_1^2p_2^2=256$. As illustrated, our proposed method shows a better performance in terms of finding the direction of the propagation path in the mmWave channel.We show that with 64 antennas, our proposed approach achieves greater that 50 percent alignment success for SNR values down to -9 dB. Our acquisition rate is 10 percent greater than RdPerm, which in a practical scenario results in a 10 percent lower need for training retransmission.

\begin{figure}[t]
\centering
 \includegraphics[scale=.6]{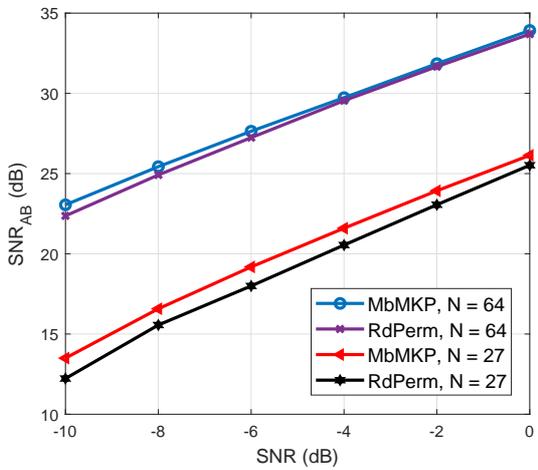}
  \caption{SNR$_\text{AB}$ vs SNR for $L=1$.}\label{effective64}
\end{figure}

Also, as is illustrated in Fig.~\ref{effective64}, our proposed approach shows a superior performance compared to RdPerm in terms of the SNR after BF. For example, when SNR $=$ -10 dB for the case with $N=27$, our approach outperforms RdPerm by more than 1 dB. Note that the number of measurements for the scenarios with $N=27$ and $N=64$ in Fig.~\ref{effective64} are the same as those of Fig.~\ref{OneCluster64ant}.
\section{Conclusion}\label{conclusion}
In this paper, we have proposed a new deterministic sensing matrix for the beam alignment problem in mmWave systems. Our proposed sensing matrix is sparse, which is computationally efficient. We have shown that our proposed approach meets the restricted isometry property. Based on the proposed deterministic sensing matrix, we have designed the BF vectors needed to probe the channel in the training step. Simulation results verify that our proposed approach outperforms the method employing the random BF technique.

\end{document}